\documentclass{iau}
\usepackage{graphicx}
\usepackage{natbib}
\usepackage{amsmath}

\usepackage[T1]{fontenc}
\usepackage{aecompl}

\usepackage[colorlinks, linkcolor={red}, citecolor={blue}, urlcolor={blue}]{hyperref}

\newcommand{\ebv}{\ensuremath{E(B-V)}}		
\newcommand{\ebvm}{\ensuremath{E(B-V)_{\text{M33}}}}		
\newcommand{\ebvmw}{\ensuremath{E(B-V)_{\text{MW}}}}		
\newcommand{\nai}{Na~\textsc{i}}		

\title[Diffuse interstellar bands in M33] 
{Diffuse interstellar bands in M33}

\author[K. T. Smith \etal]   
{Keith T. Smith$^{1,2}$\footnote{Email: {\tt kts@ras.org.uk}}, Martin A. Cordiner$^3$, Christopher J. Evans$^4$, Nick~L.~J.~Cox$^5$ \and Peter J. Sarre$^1$}

\affiliation{$^1$School of Chemistry, The University of Nottingham, Nottingham, NG7 2RD, UK\\[\affilskip]
$^2$Royal Astronomical Society, Burlington House, Piccadilly, London, W1J 0BQ, UK\\[\affilskip]
$^3$NASA Goddard Space Flight Center, 8800 Greenbelt Road, Greenbelt, MD 20770, USA\\[\affilskip]
$^4$UK Astronomy Technology Centre, Royal Observatory Edinburgh, Edinburgh, EH9 3HJ, UK\\[\affilskip]
$^5$Institute for Astronomy, K.U. Leuven, Celestijnenlaan 200D, bus 2401, Leuven, Belgium}

\pubyear{2013}
\volume{297}  
\pagerange{xxx--yyy}
\date{?? and in revised form ??}
\jname{The Diffuse Interstellar Bands}
\editors{J. Cami \& N. L. J. Cox, eds.}
\begin{document}

\maketitle

\begin{abstract}
We present the first sample of diffuse interstellar bands (DIBs) in the nearby galaxy M33. Studying DIBs in other galaxies allows the behaviour of the carriers to be examined under interstellar conditions which can be quite different from those of the Milky Way, and to determine which DIB properties can be used as reliable probes of extragalactic interstellar media. Multi-object spectroscopy of 43 stars in M33 has been performed using Keck/DEIMOS. The stellar spectral types were determined and combined with literature photometry to determine the M33 reddenings \ebvm. Equivalent widths or upper limits have been measured for the $\lambda$5780 DIB towards each star. DIBs were detected towards 20 stars, demonstrating that their carriers are abundant in M33. The relationship with reddening is found to be at the upper end of the range observed in the Milky Way. The line of sight towards one star has an unusually strong ratio of DIB equivalent width to \ebvm, and a total of seven DIBs were detected towards this star.
\end{abstract}

\firstsection 
\section{Introduction}
Although much is known about the diffuse interstellar bands (DIBs), almost all information on the bands and their carriers has been derived by studying the interstellar medium (ISM) within a few kiloparsecs of the Sun. These observations probe the behaviour of the carriers when subjected to the conditions present in the Milky Way (MW) ISM. However, the carriers may behave differently when exposed to different conditions, which may provide useful information on the carriers. The ISM in other galaxies can have quite different properties, for example the metallicity, ambient ultraviolet (UV) field, gas-to-dust ratio etc.

Observations of DIBs in other environments can be used to test known correlations e.g. between DIB equivalent width and \ebv, $N(\text{H\,\textsc{i}})$, UV field etc. If these correlations are found to vary, this will provide information on the carriers. If instead the correlations are found to be universal, then observations of DIBs can be used as proxies for other interstellar conditions which may be much more difficult to determine in extragalactic objects.

The study of DIBs in extragalactic environments has expanded in recent years with observations of the Large and Small Magellanic Clouds \citep{Welty2006,Cox2006,Cox2007}, supernovae \citep{Cox2008,Thoene2009} and M31 \citep{Cordiner2008m31,Cordiner2011}. A review of extragalactic DIB detections and their utility is given by Cordiner (2013, this volume).

M33 is an attractive target for studies of extragalactic DIBs because:
\begin{itemize}
 \item It is sufficiently close that individual stars can be resolved in seeing limited observations
 \item M33 is a late-type spiral with many early-type stars, providing suitable targets
 \item The low inclination of the disk means the stars are seen through a relatively simple interstellar column (unlike e.g. M31)
 \item Low foreground Milky Way extinction \ebvmw
 \item There is a radial velocity offset between material in M33 and the MW foreground, allowing the location of the DIBs and atomic lines to be determined from their velocities
 \item M33 has an angular size of $\sim30'\times35'$, which is well matched to the fields of view of many multi-object spectrographs
 \item A large body of multiwavelength data is available for comparison purposes
\end{itemize}
We therefore embarked upon a project to observe DIBs in M33. The first detections in this galaxy were reported towards a single star by \citet{Cordiner2008m33}; here we present an expanded sample of forty-three stars.

\section{Observations}
\label{sec:obs}
\begin{figure}
 \includegraphics[width=\textwidth]{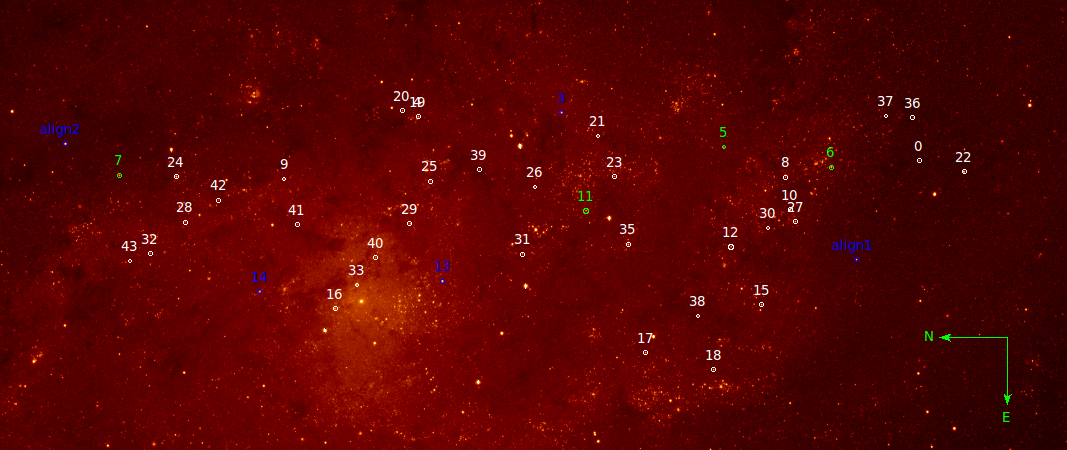}
 \caption{Locations of the target stars in M33. The background is a logarithmically scaled $V$ band image from \citet{Massey2006}; targets are circled and numbered. Stars 3, 13 and 14 were eliminated due to unsuitable spectral types (see section~\ref{sec:types}), stars 5, 6, 7 and 11 were discarded due to blended photometry (see Fig.~\ref{fig:blends}), and align1 and align2 are foreground MW stars used to align the slitmask.}
 \label{fig:targets}
\end{figure}

A total of forty-three stars in M33 were observed using the Deep Imaging Multi-Object Spectrograph [DEIMOS, \citet{Faber2003}] on the 10\,m Keck~II telescope. The observations were performed on the night of 2003~November~1 as part of a stellar metallicity study. The targets were photometrically selected to be bright ($V<19$) A- and B-type supergiants; their distribution is shown in Figure~\ref{fig:targets}. Using two different tilts of the 1200G grating, the observations covered wavelengths of approximately 3,500--9,000\,\AA\ (with two small gaps). The resolution was constant at $\Delta\lambda\approx1.7$\,\AA, which corresponds to resolving powers of $R\equiv\frac{\lambda}{\Delta\lambda}\sim2,200$--$5,000$. The continuum signal-to-noise ratio at 6,000\,\AA\ ranged from 30--300, with most stars in the range 60--100. Full details of the observations and data reduction are given in \citet{Smith2010} and Smith \etal\ (in preparation).

\section{Results}

\subsection{Spectral types and reddenings}
\label{sec:types}

The spectral types of each target were determined by comparison to the standard stars of \citet{Evans2003} and \citet{Evans2004}, with luminosity classes assigned based on the H~$\gamma$ equivalent width \citep{Evans2004}. We subsequently discarded 2 foreground MW K-type stars used for alignment, and 1 Wolf-Rayet and 3 luminous blue variables (LBVs) in M33 as unsuitable for the study of DIBs [the LBV spectra have been published in \citet{Clark2012}]. This left 1 O-type, 27 B-type, and 9 later-type stars in M33, mostly supergiants.

\begin{figure}
 \centering
 \includegraphics[width=0.32\textwidth]{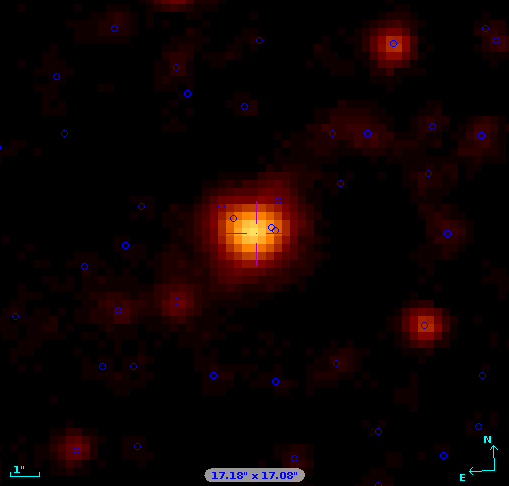} \includegraphics[width=0.32\textwidth]{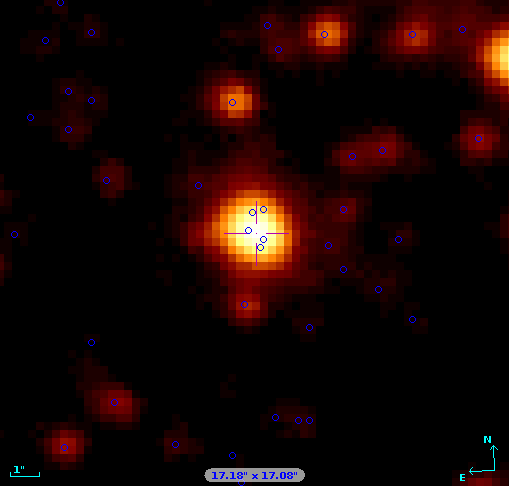} \includegraphics[width=0.32\textwidth]{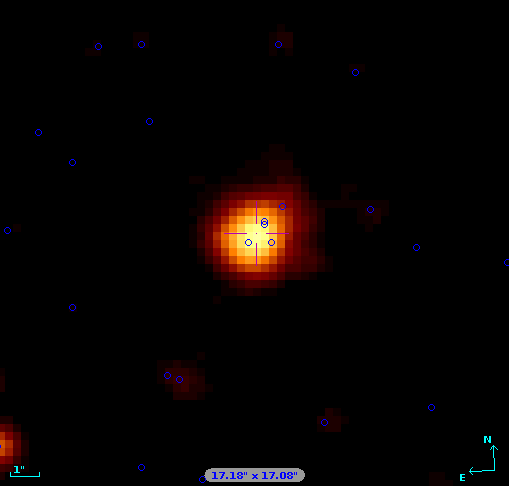}
 \caption{Three stars which were discarded due to blended photometry. The images are logarithmically scaled $V$ band and the circles show the catalogued sources, both from \citet{Massey2006}.}
 \label{fig:blends}
\end{figure}

Total reddenings \ebv\ were determined from the photometry of \citet{Massey2006} and the intrinsic colours of \citet{Fitzgerald1970} and \citet{Martins2006}. A further four targets were rejected as having blended photometry in \citet{Massey2006}, see Figure~\ref{fig:blends}. The foreground MW reddening in this direction of $\ebvmw=0.04\pm0.02$ \citep[see][]{Cordiner2008m33,Smith2010} was subtracted to give the reddening in M33, \ebvm.

\subsection{DIB detections}

\begin{figure}
 \includegraphics[width=\textwidth]{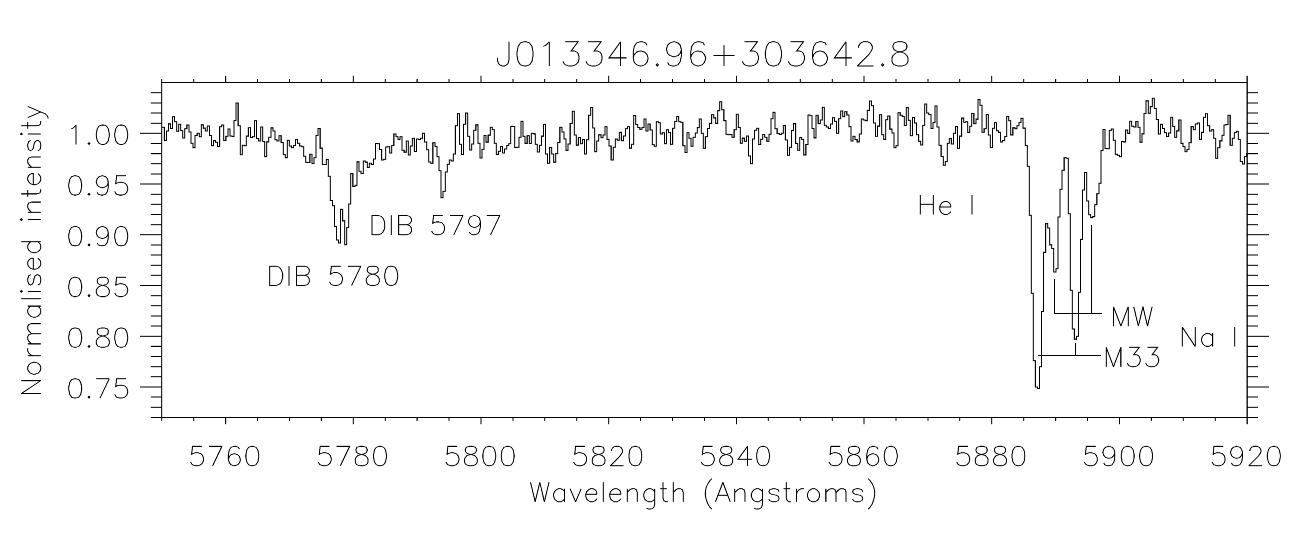}
 \caption{Spectrum of J013346.96+303642.8 covering the wavelengths of the $\lambda5780$ \& $\lambda5797$ DIBs and the \nai~D~lines. Both DIBs are clearly detected, and separate M33 and foreground MW components can be seen in the \nai\ doublet. The stellar He~\textsc{i} D$_3$ line is also detected.}
 \label{fig:example}
\end{figure}

We analysed the \nai\ D lines to determine the velocities of the ISM absorption in M33 and the MW foreground (e.g. Figure~\ref{fig:example}), then searched for DIBs at the M33 velocity. Equivalent widths were measured using standard DIB profiles, shifted and scaled to fit the observations using a $\chi^2$ minimisation routine. The standard profiles were derived from a high resolution, high signal-to-noise spectrum of $\beta^1$~Sco \citep{Cordiner2006}, degraded to the DEIMOS resolution. This could cause systematic errors if the DIBs in M33 have different profile shapes to those towards $\beta^1$~Sco, but we see no evidence for any differences. See \citet{Cordiner2011} for details of this procedure, which reduces systematic errors when measuring broad, asymmetric DIBs in low signal-to-noise spectra. DIBs were detected towards a total of 20 stars.

\begin{figure}
 \includegraphics[width=\textwidth]{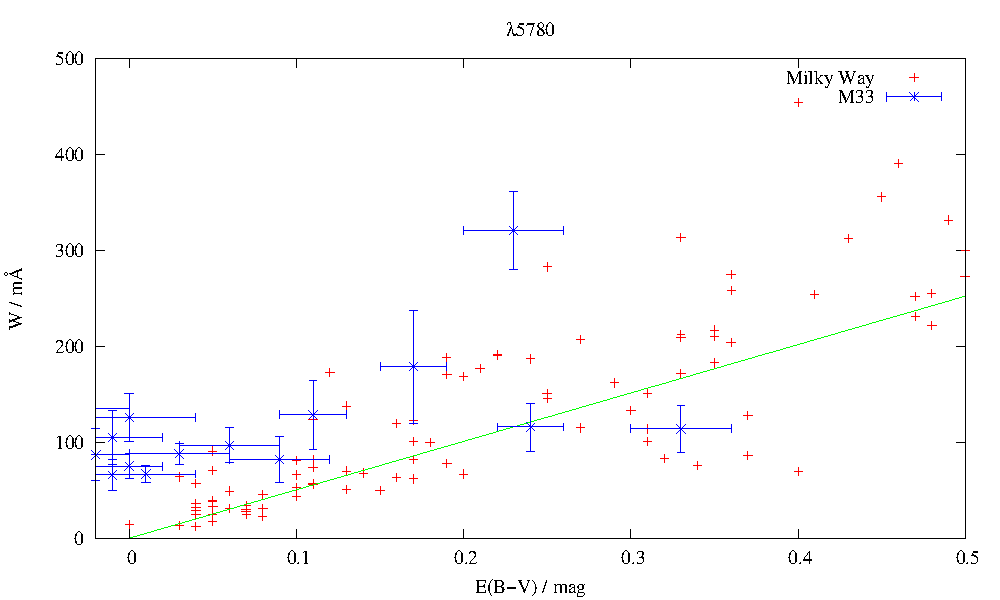}
 \caption{Relationship between the equivalent width $W$ of the $\lambda5780$ DIB and the reddening in M33 \ebvm. Milky Way comparison data, and the line showing the mean MW relation, are taken from \citet{Friedman2011}. The M33 detections lie towards the upper end of the range seen in the MW.}
 \label{fig:5780_reddening}
\end{figure}

The relationship between the equivalent width $W$ of the $\lambda5780$ DIB and \ebvm\ is shown in Figure~\ref{fig:5780_reddening}. Most of the measurements in M33 lie towards the upper limit of the range seen in the MW, particularly at low reddenings. However, this might just be a selection effect, because equivalent widths of $\lesssim80$\,m\AA\ would not be reliably detected at the signal-to-noise of our spectra. There could be contamination of the measurements by the $\lambda5778$ DIB, which is too weak and shallow to reliably detect in our spectra but frequently overlaps $\lambda5780$ in observations of targets in the MW. The measurements also cover only a restricted range in \ebvm. Analysis of the $\lambda6283$ DIB and comparison with data at other wavelengths (e.g. 8\,$\mu$m PAH emission) is under way.

One star, J013346.96+303642.8, has especially strong DIBs for its reddening and is the highest equivalent width detection in Figure~\ref{fig:5780_reddening}. A total of 7 DIBs were detected towards this star, at velocities corresponding with the M33 \nai\ absorption: $\lambda\lambda5705$, 5780, 5797, 6203, 6269, 6283 and 6613, plus a tentative detection of $\lambda4428$. A full analysis of the spectrum of this star was presented in \citet{Cordiner2008m33}.

\section{Summary}

We have presented the first sample of diffuse interstellar bands in the nearby galaxy M33. Forty-three stars were observed using the DEIMOS spectrograph on Keck. Stellar spectral types were assigned for all stars, and M33 reddenings \ebvm\ derived. DIBs were detected towards 20 stars, demonstrating that their carriers are abundant in this galaxy. The relationship between the equivalent width of the $\lambda5780$ DIB and \ebvm\ was found to be at the upper limit of the range seen in the Milky Way, although this may be a selection effect whereby weak DIBs are not detected in our spectra. One star, J013346.96+303642.8, has especially strong DIBs for its reddening and seven different DIBs were detected towards this star.

Full results will be presented by Smith \etal\ (in preparation).

\section*{Acknowledgements}

KTS acknowledges travel grants from the IAU and the Royal Astronomical Society to attend this meeting, and funding from EPSRC. The authors thank Fabio Bresolin and Norbert Przybilla for kindly providing their raw DEIMOS data.

The data presented in this paper were obtained at the W.M. Keck Observatory, which is operated as a scientific partnership among the California Institute of Technology, the University of California and the National Aeronautics and Space Administration. The Observatory was made possible by the generous financial support of the W.M. Keck Foundation.

\begin{discussion}

    \discuss{Cox}{Is there a trend with galactocentric radius?}

    \discuss{Smith}{Simply plotting which stars did/didn't have $\lambda5780$ detections did not show an obvious relation, but I haven't yet done the statistics or looked at the DIB/reddening ratio. It will be interesting to see.}

    \discuss{Sarre}{Is there a spatial correlation between the DIBs and emission from PAHs?}

    \discuss{Smith}{We're going to search for this using publicly available \textit{Spitzer} 8\,$\mu$m imaging data. We didn't see any relationship in our studies on M31, but there could be one in M33.}

    \discuss{Foing}{The line of sight samples gas in both the halo and disk of the target galaxy. Can these be separated in an inclined system, where the two will have different velocities?}

    \discuss{Smith}{In principle, yes. But the velocity offsets between the disk and halo are small, even in a more inclined system like M31. We don't see any evidence for DIBs in the Milky Way halo gas, so presumably the vast majority of any absorption arises in the disks of each galaxy.}

\end{discussion} 

\end{document}